{
}

\newcommand{\putFig}[3]{
        \begin{figure}[htbp]
		\centering 
		\includegraphics[width=#3in]{#1}
                \caption{#2}
                \label{fig:#1}
        \end{figure} }
        
\newcommand{\putFrag}[4]{
        \begin{figure}[htbp]
		\centering
		  #4
 		\includegraphics[width=#3in]{#1} 
		  \caption{#2}
                \label{fig:#1}
        \end{figure} }

\newcommand{\E}{\text{E}}

\renewcommand{\vec}[1]{{\ensuremath{\boldsymbol{#1}}}}

\newtheorem{thm}{Theorem}

\newcommand{\of}[1]{^{\scriptscriptstyle{(#1)}}}

{}
{}
{}
{}
{}
{}
{}
{}
{}
{}

\documentclass{IEEEtran}
\usepackage{setspace,amsfonts,graphicx,cite,bm,psfrag,amsmath,amssymb}

\begin{document}
\title{Non-Hierarchical Clock Synchronization for Wireless Sensor Networks}
\author{D.~Richard Brown III, Andrew G.~Klein, and Rui Wang
\thanks{Worcester Polytechnic Institute, 100 Institute Road, Worcester, MA 01609, \{drb,klein,rwang\}@wpi.edu}}

\newcommand{\ee}{\vec{e}}

\maketitle
\setcounter{page}{1} \thispagestyle{plain}

\hyphenation{time-stamped}
\renewcommand{\ss}{stepsize }

\begin{abstract}
Time synchronization is important for a variety of applications in wireless sensor networks including scheduling communication resources, coordinating sensor wake/sleep cycles, and aligning signals for distributed transmission/reception. This paper describes a non-hierarchical approach to time synchronization in wireless sensor networks that has  low overhead and can be implemented at the physical and/or MAC layers. Unlike most of the prior approaches, the approach described in this paper allows all nodes to use exactly the same distributed algorithm and does not require local averaging of measurements from other nodes. Analytical results show that the non-hierarchical approach can provide monotonic expected convergence of both drifts and offsets under broad conditions on the network topology and local clock update stepsize. Numerical results are also presented verifying the analysis under two particular network topologies. 
\end{abstract}

\begin{IEEEkeywords}
synchronization, timing, wireless sensor networks, consensus clock 
\end{IEEEkeywords}


\section{Introduction}

Synchronization is the process of establishing a common notion of time among two or more entities. In the context of wired and wireless communication networks, synchronization enables coordination among the nodes in the network and can facilitate scheduling of communication resources, interference avoidance, event detection/ordering, data fusion, and coordinated wake/sleep cycles \cite{Maggs:SJ:2012}. Precise synchronization, to the order of a fraction of a carrier period, can also enable efficient distributed transmission schemes such as retrodirective distributed beamforming \cite{Preuss:TSP:2011}.


Standardized protocols for synchronizing devices in a network include Network Time Protocol (NTP) \cite{Mills:TCOM91} and Precision Time Protocol (PTP, also known as IEEE 1588) \cite{IEEE1588}. Both NTP and PTP are hierarchical synchronization systems: devices in the network are assigned to a class or {\em stratum} and those with the lowest stratum number are assumed to be perfectly synchronized with Coordinated Universal Time (UTC).  Nodes with higher stratum numbers synchronize their clocks via TCP/IP messages with nodes having lower stratum numbers. While NTP is currently responsible for synchronizing the clocks of most of the devices connected to the internet, it (and PTP) are generally considered too cumbersome for networks of inexpensive computationally-constrained sensor nodes.

Several synchronization protocols have been developed to address some of the shortcomings of NTP in a sensor-network scenario. Protocols based on bidirectional messaging between nodes in adjacent levels of a hierarchical tree include the timing-sync protocol for sensor networks (TSPN) \cite{Ganeriwal:ACMSENSYS:2003}, lightweight tree-based synchronization (LTS) \cite{VanGreunen:WSNA:2003}, and tiny-sync/mini-sync \cite{Sichitiu:WCNC:2003}. While the details of these protocols differ, they are all based on the establishment of a hierarchical tree structure and the goal is to synchronize every node in the network to the root node. The use of bidirectional messages allows for disambiguation of propagation delay from clock offset, making these protocols potentially more accurate than synchronization schemes based on unidirectional messages.

The flooding time synchronization protocol (FTSP) \cite{Maroti:ENSS:2004}, reference broadcast synchronization (RBS) \cite{Elson:ACMSIGOPS:2002}, and CesiumSpray \cite{Verissimo:JRTS:1997} are among the most widely cited examples of synchronization protocols for sensor networks based on unidirectional messaging. While the use of unidirectional messages potentially reduces overhead, it prevents propagation delay compensation and limits the attainable synchronization accuracy. FTSP specifies a hierarchical structure with timestamped messages flowing from the root node to nodes in lower levels of the tree. RBS and CesiumSpray are based on broadcast beacon transmissions followed by the exchange of timestamps among receivers in the broadcast range of the beacon node. In networks with more than one beacon node, RBS gateway nodes are specified to transform timestamps between broadcast domains. While RBS and Cesium Spray are non-hierarchical, they still require some network structure (and corresponding overhead) in the establishment of special beacon nodes, the establishment of the sets of client nodes within the broadcast range of each beacon node, and the exchange of timestamps among the set of client nodes as pointed out in \cite{Sivrikaya:NETWORK:2004}. 

A handful of synchronization techniques based on consensus and diffusion have also recently been proposed. While some of these techniques require network hierarchy, e.g.~\cite{Su:TAN:2005,Li:TC:2006}, others are non-hierarchical and use local averaging of clock values reported by neighboring nodes to achieve consensus \cite{Li:TC:2006, Solis:CDC:2006, Schenato:CDC:2007, Maggs:SJ:2012}. All of these techniques are based on MAC layer timestamping. Physical layer synchronization techniques have also recently been reported including carrier synchronization for coordinated cellular base-station transmissions \cite{Preuss:TSP:2011}, cooperative synchronization of pulse-coupled clocks via spatial averaging \cite{Hu:TIT:2006}, consensus synchronization of clock drifts from carrier frequency measurements \cite{Rahman:SSP:2012}, and carrier-phase based pairwise ranging and synchronization \cite{Bidigare:SSP:2012}. Experimental results in \cite{Bidigare:SSP:2012} show carrier-phase techniques can achieve offset compensation to accuracies on the order of 10~ps.

This paper describes a non-hierarchical approach to ad hoc network synchronization based on random bidirectional pairwise message exchanges between nodes. Our approach is described in two steps: (i) drift compensation and (ii) offset compensation.  For clarity we present these steps as sequential tasks, but they can also be performed simultaneously using measurements at the MAC layer and/or physical layer. Since our approach is non-hierarchical and based on random bidirectional pairwise message exchanges, it does not require any coordination among the nodes and can be embedded into existing network traffic. Our approach is especially amenable to physical layer synchronization techniques which glean drift and offset measurements from existing data packets.  All of the nodes in the network can run exactly the same algorithm, irrespective of the network topology. Since our non-hierarchical synchronization algorithm is not based on local averaging, the overhead of our approach is potentially less than the consensus techniques described in \cite{Li:TC:2006, Solis:CDC:2006, Schenato:CDC:2007, Maggs:SJ:2012}. We provide analysis proving non-hierarchical synchronization exhibits monotonic expected convergence under broad conditions on the network topology and clock update stepsize. Numerical examples showing convergence and divergence of the proposed non-hierarchical network synchronization technique under different network topologies are also provided.

The rest of the paper is organized as follows.  We first introduce a probabilistic messaging model and the relevant local clock parameters in Section~\ref{sec:model}. Then, we describe a non-hierarchical network synchronization technique based on random pairwise exchanges in Section~\ref{sec:syncmethod}. Convergence results are also provided in this section. Numerical results are given in Section~\ref{sec:sims}, followed by conclusions in Section~\ref{sec:conclusions}. Proofs of the theorems are given in the Appendices.

\underline{Notation}: Vectors and matrices are denoted by boldface letters. $\vec{I}_N$ denotes the $N \times N$ identity matrix. The vectors $\vec{1}_N$, $\vec{0}_N$, and $\ee_N$ denote a vector of all ones, a vector of all zeros, and a vector of all zeros with a one in the last position respectively, all in $\mathbb{R}^N$. $\left\|\cdot\right\|$ represents the Euclidean norm of the enclosed vector. We use ${\rm E}\left\{\cdot\right\}$ and $\left(\cdot\right)^T$ for expectation and transposition.

\section{System Model}
\label{sec:model}

We assume a time-division multiplexed network of $N$ nodes with transmit/receive topology specified by a probability matrix $\vec{P}$ with $i,j^{\rm th}$ entry $p_{i,j}$ corresponding to the probability that node~$i$ initiates an exchange of messages with node~$j$ at any particular instant in time. The case $p_{i,j} = p_{j,i} = 0$ corresponds to the situation where node~$i$ and node~$j$ do not communicate. Note that $p_{i,i} = 0$ for all $i=1,\dots,N$ and $\sum_i \sum_j p_{i,j} = 1$. We do not necessarily assume $p_{i,j} = p_{j,i}$; for example, $p_{i,j}>0$ and $p_{j,i}=0$ corresponds to the case where $i$ initiates message exchanges with node~$j$ but node $j$ never initiates message exchanges with node~$i$.

Since all of the channels in the system are time-division duplexed (TDD), we assume reciprocal propagation delays $\psi_{i,j} = \psi_{j,i}$ in each link. Basic electromagnetic principles have long established that channel reciprocity holds at the antennas when the channel is accessed at the same frequency in both directions \cite{Fettweis:VTMAG:2006}. Channel reciprocity can also be quite accurate at intermediate-frequency (IF) and/or baseband if a reciprocal transceiver architecture is used \cite{Parish:Patent:1997} and can be further improved through transceiver calibration techniques to remove I/Q imbalance effects \cite{Bourdoux:RAWCON:2003,Guillaud:ISSPA:2005}.

The nodes in the network do not possess a common notion of time. The following section presents a model of local and reference time that will be subsequently used in the description and analysis of the pairwise synchronization protocol.

\subsection{Reference Time and Local Time}
The focus of this paper is the description and analysis of a synchronization technique for devices in a wireless ad-hoc network. To support this focus, it is necessary to explicitly present a model of local time at each node and describe how the local time at each node relates to a notion of ``reference'' time. The notation $t$ refers to the reference time, i.e.~the ``true'' time, in the system. All time-based quantities such as propagation delays and/or frequencies are specified in reference time unless otherwise noted. 

None of the nodes have knowledge of the reference time $t$. The local time at node~$i$ modeled as
\begin{equation*}
  t_i = t + \Delta_i(t)
\end{equation*}
where $\Delta_i(t)$ is a non-stationary random process that captures the effect of clock drift, fixed local time offset, local oscillator phase noise, and frequency instability \cite{Baghdady:PROC:1965}. Over short time periods, a reasonable first-order model of local time can be written as
\begin{equation*}
  t_i = \beta_i t + \Delta_i
\end{equation*}
where $\beta_i$ represents the nominal relative rate of the clock at node~$i$ with respect to the reference time and $\Delta_i$ is the local clock offset at $t=0$. None of the nodes in the network have knowledge of $\beta_i$ or $\Delta_i$.

\section{Non-Hierarchical Network Synchronization}
\label{sec:syncmethod}
This section describes a non-hierarchical technique for network synchronization that allows each node in the network to arrive at a common clock drift $\beta_i$ and clock offset $\Delta_i$. The goal is not to force $\beta_i=1$ and $\Delta_i=0$. Rather, as is often the case in ad hoc network synchronization \cite{Maggs:SJ:2012}, the goal is to drive the clock drifts and offsets to {\em common} values $\bar{\beta}$ and $\bar{\Delta}$ across the network. For conceptual simplicity, we describe our non-hierarchical approach to network synchronization as a two-step process: (i) drift compensation and (ii) offset compensation. In practice, both drift and offset compensation can be performed simultaneously, since pairwise drift estimates can be inferred ``for free'' from normal network traffic.

\subsection{Step 1: Drift Compensation}

Since the reference time $t$ is unobservable, this section develops a drift compensation framework in the context of {\em pairwise} drift estimates and compensation. In timeslot $k$, the pairwise clock drift between node~$i$ and node~$j$ is defined as
\begin{equation*}
  \beta_{i,j}[k] = \frac{d}{dt} (t_i-t_j) = \beta_i[k]-\beta_j[k].
\end{equation*}
There are $N(N-1)$ such pairwise drifts in the network. Since $\beta_{j,i}[k]=-\beta_{i,j}[k]$, we can define the network pairwise drift vector at time $k$ as 
\begin{equation}
  \label{eq:beta}
  \vec{\beta}[k] = \left[ \begin{matrix} \vec{\beta}_{2}[k] \\  \vec{\beta}_{3}[k] \\ \vdots \\ \vec{\beta}_{N}[k] \end{matrix} \right] \in \mathbb{R}^{N(N-1)/2}
\end{equation}
where $\vec{\beta}_j[k] = [\beta_{1,j}[k],\dots,\beta_{j-1,j}[k]]^\top \in \mathbb{R}^{j-1}$ is a vector of pairwise drifts $\beta_{i,j}$ with respect to node $j$ for all $i<j$. 
Furthermore, since $\beta_{i,j} = \beta_{i,N}-\beta_{j,N}$ for all $i,j$ pairs, we can write
\begin{equation}
  \label{eq:qimplicit}
  \vec{\beta}[k] = \vec{Q}_N \vec{\beta}_N[k]
\end{equation}
where $\vec{Q}_N \in \mathbb{R}^{N(N-1)/2 \times N-1}$. As a specific example, in the $N=4$ node case we have
\begin{equation*}
  \vec{\beta}[k] = \left[ \begin{matrix}
  \beta_{1,2}[k] \\
  \beta_{1,3}[k] \\
  \beta_{2,3}[k] \\
  \beta_{1,4}[k] \\
  \beta_{2,4}[k] \\
  \beta_{3,4}[k]
  \end{matrix}
  \right] = 
  \left[ \begin{matrix}
    1 & -1 & 0\\
    1 & 0 & -1 \\
    0 & 1 & -1 \\
    1 & 0 & 0\\
    0 & 1 & 0 \\
    0 & 0 & 1 
    \end{matrix}
  \right]
  \left[ \begin{matrix}
  \beta_{1,4}[k] \\
  \beta_{2,4}[k] \\
  \beta_{3,4}[k]   
  \end{matrix}
  \right] = \vec{Q}_4 \vec{\beta}_4[k].
\end{equation*}
In general, $\vec{Q}_N$ has elements equal to $-1$, $0$, or $1$. Appendix~\ref{app:qproperties} provides a recursive definition and lists several relevant properties of $\vec{Q}_N$.

Since nodes derive their symbol rate and carrier frequency from the same local oscillator that drives the local clock, any message between a pair of nodes in the network allows for the estimation of pairwise clock drift at the physical layer through carrier frequency or symbol rate offset estimation. Pairwise clock drift can also be estimated at the MAC layer through observing multiple timestamped messages from another node in the network. Without restricting ourselves to a particular method for pairwise drift estimation, we simply assume that when the event occurs that nodes~$i$~and~$j$ form a sender/receiver pair, node $i$ adjusts its local clock rate based on its local estimate of its pairwise drift with node~$j$. More precisely, node~$i$ forms the estimate $\hat{\beta}_{j,i}[k]$ and adjusts its local clock drift
\begin{equation}
  \label{eq:localdriftupdate}
  \beta_{i}[k+1] = \beta_{i}[k] + \mu \hat{\beta}_{j,i}[k]
\end{equation}
where $\mu > 0$ is a \ss parameter. All other local clock drifts in the network (including node~$j$, the receiver node) are not updated, i.e.
\begin{equation*}
  \beta_{\ell}[k+1] = \beta_{\ell}[k] \qquad \forall \ell \neq i.
\end{equation*}

Note that the local drift adjustment at node~$i$ affects the pairwise drifts $\beta_{j,i}$ and $\beta_{i,j}$ for all $j$. Hence, we can represent the update to the pairwise drift vector as
\begin{equation}
  \label{eq:driftupdate}
  \vec{\beta}[k+1] = \vec{\beta}[k] + \mu \vec{b}_{i,j}[k]
\end{equation} 
where $\vec{b}_{i,j}[k]$ denotes the pairwise drift adjustment vector caused by (\ref{eq:localdriftupdate}). As a specific example, suppose $N=4$ and $(i,j)=(2,3)$. The pairwise drift adjustment vector that occurs when node~2 updates its local drift can be written as
\begin{equation}
  \label{eq:driftadjustmentvector}
  \vec{b}_{2,3}[k] = \hat{\beta}_{3,2}[k] \left[-1,0,1,0,1,0\right]^\top = \hat{\beta}_{3,2}[k] \bar{\vec{q}}_2 
\end{equation}
where $\bar{\vec{q}}_i$ is the $i^{\rm th}$ column of $\bar{\vec{Q}}_N$ for $i=1,\dots,N$ with 
\begin{equation*}
  \bar{\vec{Q}}_N = \left[  \begin{matrix} \vec{Q}_N & \vec{q}_N \end{matrix} \right] \in \mathbb{R}^{N(N-1)/2 \times N}
\end{equation*} 
where $\vec{q}_N = [0,\dots,0,-1,\dots,-1]^\top$ is a vector of zeros with the last $N-1$ elements equal to $-1$. 


\subsection{Step 2: Offset Compensation}
While compensating for oscillator drift {\em syntonizes} the nodes in the network, it is not sufficient for {\em synchronization} because the fixed clock offsets among the nodes in the network are not corrected. To correct these offsets in a pairwise manner, we assume that the pairwise drift between nodes~$i$~and~$j$ is negligible and that node~$i$ and node~$j$ form a sender/receiver pair and exchange messages such that node~$i$ can disambiguate its pairwise clock offset with node~$j$ from the propagation delay $\psi_{i,j}=\psi_{j,i}$. As one example of how this can be achieved \cite{Ganeriwal:ACMSENSYS:2003}, consider the exchange of timestamped packets as shown in Figure~\ref{fig:sender_receiver} below. Given a packet transmitted by node~$i$ in local time $t\of{a}_i$, it arrives at node~$j$ in local time $t\of{b}_j = t\of{a}_i + \psi_{i,j}+\Delta_j-\Delta_i$. The response from node~$j$ contains the local timestamps $t\of{b}_j$ and $t\of{c}_j$ and arrives at node~$i$ at local time $t\of{d}_i = t\of{c}_j + \psi_{i,j}+\Delta_i-\Delta_j$. After receiving the response, node~$i$ can compute the pairwise clock offset to node~$j$ as 
\begin{align*}
  \frac{(t\of{b}_j-t\of{a}_i)-(t\of{d}_i-t\of{c}_j)}{2} 
  &={} \Delta_j-\Delta_i = \Delta_{j,i}.
\end{align*}
After an exchange of messages with node~$j$, node~$i$ forms the estimate $\hat{\Delta}_{j,i}[k]$ and adjusts its local clock offset
\begin{equation}
  \label{eq:localoffsetupdate}
  \Delta_{i}[k+1] = \Delta_{i}[k] + \mu \hat{\Delta}_{j,i}[k]
\end{equation}
where $\mu > 0$ is a \ss parameter. All other local clock offsets in the network are not updated. We can represent the update to the pairwise offset vector $\vec{\Delta}[k] \in \mathbb{R}^{N(N-1)/2}$ as
\begin{equation*}
  \vec{\Delta}[k+1] = \vec{\Delta}[k] + \mu \vec{d}_{i,j}[k]
\end{equation*} 
where $\vec{d}_{i,j}[k]$ denotes the pairwise offset adjustment vector caused by (\ref{eq:localoffsetupdate}). 

\putFrag{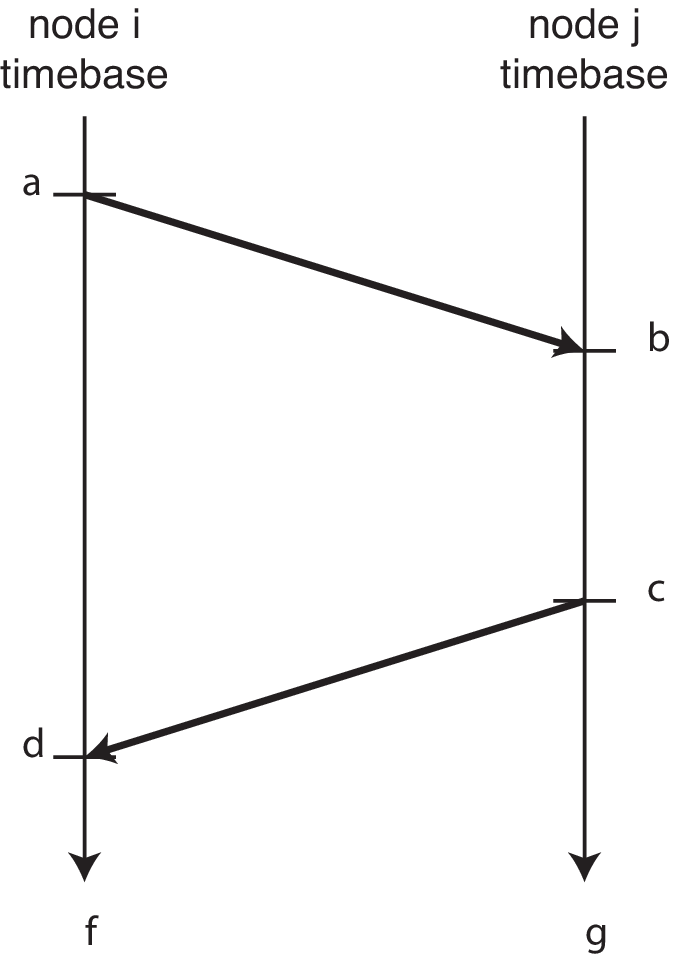}{Sender/receiver two-way message exchange.}{1.6}{
    \psfrag{a}[r][Bl][1]{$t\of{a}_i$}
    \psfrag{b}[l][Bl][1]{$t\of{b}_j$}
    \psfrag{c}[l][Bl][1]{$t\of{c}_j$}
    \psfrag{d}[r][Bl][1]{$t\of{d}_i$}   
    \psfrag{f}[][l][1]{$t_1=t+\Delta_1$}
    \psfrag{g}[][l][1]{$t_2=t+\Delta_2$}  
} 

In this context, offset compensation is conceptually identical to drift compensation as discussed in the previous section. The pairwise offsets and corresponding update vectors have the same form as the pairwise drift updates in~(\ref{eq:driftadjustmentvector}). 

\subsection{Convergence Analysis}
\label{sec:convergence}
We use $\| \vec{\beta}[k] \|^2_2$ and $\| \vec{\Delta}[k] \|^2_2$ as a measure of the overall network pairwise drift and offset alignment at time $k$, respectively, with smaller values corresponding to better overall pairwise alignment. Since pairwise offset updates occur in the same manner as pairwise drift updates, our analysis here will focus on convergence of the pairwise drifts with the understanding that the theorems also apply to the convergence of pairwise offsets by substituting $\vec{\Delta}[k]$ for $\vec{\beta}[k]$ and $\vec{d}_{i,j}[k]$ for $\vec{b}_{i,j}[k]$.

In general, pairwise drift compensation does not necessarily improve the overall pairwise drift alignment metric. For example, suppose $\vec{\beta}[k] = [-1,-2,-1,1,2,3]^\top$ with $\| \vec{\beta}[k] \|^2_2=20$. Nodes~2~and~3 form a sender/receiver pair and node~2 adjusts its local drift by $\beta_{3,2}[k]=-\beta_{2,3}[k]$, causing the drift adjustment vector $\vec{b}_{2,3}[k] =  \left[-1,0,1,0,1,0\right]^\top$. With $\mu=1$, the resulting pairwise drifts at time $k+1$ can be written as $\vec{\beta}[k+1] = [-2,-2,0,1,3,3]^\top$ with $\| \vec{\beta}[k+1] \|^2_2=27$. In this example, even though node~2 has perfectly corrected its drift with respect to node~3, the overall network drift alignment metric has become worse. Under certain conditions on $\mu$ and $\vec{P}$, however, we will show that the drift update prescribed in (\ref{eq:driftupdate}) with perfect estimates $\hat{\beta}_{j,i}[k] = \beta_{j,i}[k] = -\beta_{i,j}[k]$ results in {\em monotonic expected convergence} such that $\E \left\{ \|\vec{\beta}[k+1]\|^2_2 \, | \, \vec{\beta}[k]\right\} < \|\vec{\beta}[k]\|^2_2$ for all $\vec{\beta}[k] \neq 0$.

Prior to stating the convergence theorems, we will define some relevant quantities. First, define the mapping
\begin{equation*}
  m(i,j) = i+\sum_{k=2}^{j-1}(k-1)
\end{equation*} for integer $i,j$ satisfying $1 \le i < j \le N$. Note that $m(i,j) \in \{1,\dots,N(N-1)/2\}$ corresponds to the position of the entry in $\vec{\beta}[k]$ containing the pairwise drift $\beta_{i,j}[k]$. 

In the case with perfect estimates such that $\hat{\beta}_{j,i}[k] = \beta_{j,i}[k] = -\beta_{i,j}[k]$, the pairwise drift adjustment vector $\vec{b}_{i,j}[k]$ is linear in $\vec{\beta}[k]$. Hence, we can write
\begin{align}
  \label{eq:linearupdate}
  \vec{b}_{i,j}[k] &={} -\bar{\vec{q}}_i \beta_{i,j}[k] = -\bar{\vec{Q}}_N \vec{E}_{i,j} \vec{\beta}[k]
\end{align}
where $\vec{E}_{i,j} \in \mathbb{R}^{N \times N(N-1)/2}$ is a matrix of all zeros except for a ${\rm sign}(j-i)$ in position $(i,m(i,j))$ if $i<j$ or position $(i,m(j,i))$ if $i>j$. The matrix $\vec{E}_{i,j}$ effectively selects $\bar{\vec{q}}_i$ from $\bar{Q}_N$ and $\beta_{i,j}[k]$ from $\vec{\beta}[k]$ when $i<j$ or $-\beta_{i,j}[k] = \beta_{j,i}[k]$ from $\vec{\beta}[k]$ when $i>j$.

From (\ref{eq:linearupdate}), we can write
\begin{align}
  \E\{\vec{b}_{i,j}[k] \, | \, \vec{\beta}[k]\} &={} \sum_{i=1}^N \sum_{j=1}^N p_{i,j} \vec{b}_{i,j}[k] \notag \\
  &={} -\bar{\vec{Q}}_N \left( \sum_{i=1}^N \sum_{j=1}^N p_{i,j} \vec{E}_{i,j} \right) \vec{\beta}[k] \notag \\
  &={} -\vec{R} \vec{\beta}[k]. 
  \label{eq:R}
\end{align}
Also note that
\begin{equation*}
  \| \vec{b}_{i,j}[k] \|_2^2 = \| \bar{\vec{q}}_i \beta_{j,i}[k] \|_2^2  = (N-1) \beta_{i,j}^2[k]
\end{equation*} 
where we used the fact that $\bar{\vec{q}}_i$ has $N-1$ elements equal to $\pm 1$ and that $\beta_{j,i}[k] = -\beta_{i,j}[k]$. It follows that
\begin{align}
  \E \left\{\| \vec{b}_{i,j}[k] \|_2^2 \, | \, \vec{\beta}[k] \right\} &={} (N-1) \sum_{i=1}^N \sum_{j=1}^N p_{i,j}  \beta_{i,j}^2[k] \notag \\
   &={} \vec{\beta}^\top[k] \vec{S} \vec{\beta}[k]
   \label{eq:S}
\end{align} 
where $\vec{S} \in \mathbb{R}^{N(N-1)/2 \times N(N-1)/2}$ is a diagonal matrix with entries given as $\vec{S}_{m(i,j),m(i,j)} = (N-1)(p_{i,j}+p_{j,i})$ for $1 \le i < j \le N$.

\begin{thm}[General Monotonic Expected Convergence]
\label{thm:generalcase}
Given $\vec{\beta}[k] \neq 0$, $\hat{\beta}_{j,i} = \beta_{j,i}$, and local drift updates specified by (\ref{eq:localdriftupdate}). Then
$\E \left\{ \|\vec{\beta}[k+1]\|^2_2 \, | \, \vec{\beta}[k]\right\} < \|\vec{\beta}[k]\|^2_2$ if and only if
\begin{equation}
 \label{eq:general}
  \vec{Q}_N^\top \left( \vec{R} + \vec{R}^\top - \mu \vec{S}\right) \vec{Q}_N
\end{equation}
is positive definite.
\end{thm}
A proof of this theorem is provided in Appendix~\ref{app:generalcase}.

Theorem~\ref{thm:generalcase} provides an implicit condition on the \ss $\mu$ sufficient for monotonic expected convergence. It is straightforward to check (\ref{eq:general}) numerically for a given \ss $\mu$. 

In networks where each node's transmission range exceeds the geographic span of the network, it is reasonable to model the network as having equiprobable sender/receiver pairs such that $p_{i,j} = \frac{1}{N(N-1)}$ for all $i\neq j$. The following theorem establishes an explicit condition on the \ss $\mu$ for this case.
\begin{thm}[Equiprobable Monotonic Expected Conv.]
\label{thm:equiprobable}
Given $\vec{\beta}[k] \neq 0$, $\hat{\beta}_{j,i} = \beta_{j,i}$, local drift updates specified by (\ref{eq:localdriftupdate}), and sender/receiver probabilities $p_{i,j} = \frac{1}{N(N-1)}$ for all $i \neq j$. Then $\E \left\{ \|\vec{\beta}[k+1]\|^2_2 \, | \, \vec{\beta}[k]\right\} < \|\vec{\beta}[k]\|^2_2$ if and only if
\begin{equation*}
  \mu < \frac{N}{N-1}.
\end{equation*}
\end{thm}
A proof of this theorem is provided in Appendix~\ref{app:equiprobable}. The proof relies on several properties of $\vec{Q}_N$ listed in Appendix~\ref{app:qproperties}. 


\subsection{Discussion}
\label{sec:discussion}
As a non-trivial example of a case where pairwise drift and pairwise offset compensation do not exhibit monotonic expected convergence for any $\mu > 0$, consider an $N=3$ node network with 
\begin{equation*}
  \vec{P} = \left[ \begin{matrix} 0 & 0 & 0.9 \\ 0 & 0 & 0.05 \\ 0.05 & 0 & 0 \end{matrix} \right].
\end{equation*}
Note that all nodes initiate message exchanges in this network, hence each node adapts its local drift/offset at time $k$ with nonzero probability. Straightforward calculations give
\begin{align*}
  \vec{Q}_N^\top \! \left( \vec{R} \! + \! \vec{R}^\top \!\! - \! \mu \vec{S}\right) \! \vec{Q}_N 
  & = {} 
  \left[ \begin{matrix} 3.7-1.9\mu & -0.9 \\ -0.9 & 0.2 - 0.1\mu \end{matrix} \right] .
\end{align*}
It can be numerically verified that one eigenvalue of this matrix is negative when $\mu = 0$ and that both eigenvalues decrease as $\mu$ increases. Hence, there does not exist a \ss $\mu>0$ such that this network satisfies the conditions for monotonic expected convergence.

In networks that admit monotonic expected convergence for some \ss $\mu>0$, it is possible to derive from (\ref{eq:condition2}) in Appendix~\ref{app:generalcase} the value of $\mu$ that provides the largest expected convergence step. Rewriting (\ref{eq:condition2}) as
\begin{equation*}
  g(\mu) \! = \! -2\mu\vec{\beta}^\top_N[k] \vec{Q}_N^\top \vec{R} \vec{Q}_N \vec{\beta}_N[k] + \mu^2 \vec{\beta}_N^\top[k] \vec{Q}_N^\top \vec{S} \vec{Q}_N \vec{\beta}_N[k] 
\end{equation*}
the value of $\mu$ that minimizes $g(\mu)$ is
\begin{equation*}
  \mu_{\rm opt} = \frac{\vec{\beta}^\top_N[k] \vec{Q}_N^\top \vec{R} \vec{Q}_N \vec{\beta}_N[k]}{ \vec{\beta}_N^\top[k]  \vec{Q}_N^\top \vec{S} \vec{Q}_N \vec{\beta}_N[k] }
\end{equation*}
which reduces to $\mu_{\rm opt} = \frac{N}{2(N-1)}$ in the equiprobable case since $\vec{Q}_N^\top \vec{R}\vec{Q}_N = \frac{1}{N-1}\vec{Q}_N^\top \vec{Q}_N$ and $\vec{Q}_N^\top\vec{SQ}_N = \frac{2}{N}\vec{Q}_N^\top \vec{Q}_N$ as shown in Appendix~\ref{app:equiprobable}. In the non-equiprobable case, the optimal \ss adapts with the current state of the pairwise drift vector. Exact calculation of $\mu_{\rm opt}$ requires knowledge of all of the pairwise offsets with respect to node $N$, hence it is of limited practical utility in the non-equiprobable case. Nevertheless, this result suggests that nodes could have a heuristic for adapting the local \ss to start at a large value and become smaller as the network becomes more closely synchronized.

Finally, although pairwise drift and offset compensation allow the nodes in an ad hoc network to achieve consensus on the drifts and offsets such that $\beta_i \rightarrow \bar{\beta}$ and $\Delta_i \rightarrow \bar{\Delta}$ for all $i$, it is worth mentioning that the technique can also be used to synchronize to an external source of reference time if one or more nodes in the network have access to reference time (via, e.g.~GPS). The nodes that have access to an external source of reference time simply do not behave as ``senders'' in the network. This forces the other nodes in the network to adapt to the reference time. Hence, if the conditions of Theorem~\ref{thm:generalcase} are satisfied, the network will exhibit monotonic expected convergence toward synchronization with an external reference time.

\section{Numerical Results}
\label{sec:sims}
The numerical results in this section assume a network with $N=10$ nodes, i.i.d.~Gaussian distributed initial clock offsets $\Delta_i[0]$ with standard deviation 5~ms, and i.i.d~Gaussian distributed initial drifts $\beta_i[0]$ with standard deviation 100~$\mu$s/iteration, for $i=1,\dots,N$. In iterations $k=0,\dots,99$, no synchronization updates occur. During this time, the pairwise drifts remain constant and the pairwise offsets tend to grow. For iterations $k=100,\dots,499$, the drift compensation algorithm runs with randomly selected sender/receiver pairs with probabilities specified by $\vec{P}$. For iterations $k=500,501,\dots$ the offset compensation algorithm runs, also with randomly selected sender/receiver pairs with probabilities specified by $\vec{P}$. In the following, we consider two network topologies: (i) a fully-connected network with equiprobable sender/receiver pairs and (ii) a partitioned network with equiprobable sender/receiver pairs in each partition and a single gateway node connecting the partitions.

\subsection{Equiprobable Sender/Receiver Pairs}
In this example, $p_{i,j} = \frac{1}{N(N-1)} = \frac{1}{90}$ for all $i\neq j$. Based on Theorem~\ref{thm:equiprobable}, both drift compensation and offset compensation will exhibit monotonic expected convergence if $\mu < \frac{10}{9}$. 

Figure~\ref{fig:n10_equiprobable_singlerealization} shows a single realization of the pairwise drift $\vec{\beta}[k]$ and pairwise offset $\vec{\Delta}[k]$ processes for $k=0,\dots,1000$ and fixed \ss $\mu=0.5$. The effect of the uncompensated drifts is evident in the first 100 iterations where we see linearly increasing pairwise offsets. In this example, the pairwise drifts and the pairwise offsets converge to values close to zero within a couple hundred iterations of when their respective compensation algorithms commence.

Figure~\ref{fig:n10_equiprobable_averagenorms_multimu} shows the pairwise synchronization metrics $\| \vec{\beta}[k] \|^2_2$ and $\| \vec{\Delta}[k] \|^2_2$, averaged over 1000 Monte-Carlo realizations. These results numerically demonstrate the monotonic expected convergence property of the synchronization algorithm for fixed stepsizes $\mu=0.1,0.5,1$. As expected from the analysis in Section~\ref{sec:discussion}, the case when $\mu=0.5$ provides the fastest convergence rate since this is the closest \ss value to $\mu_{\rm opt} = \frac{N}{2(N-1)} \approx 0.56$. The case when $\mu = 1.2 > \frac{10}{9}$ demonstrates divergence of the synchronization algorithm when the \ss is too large. 


\putFig{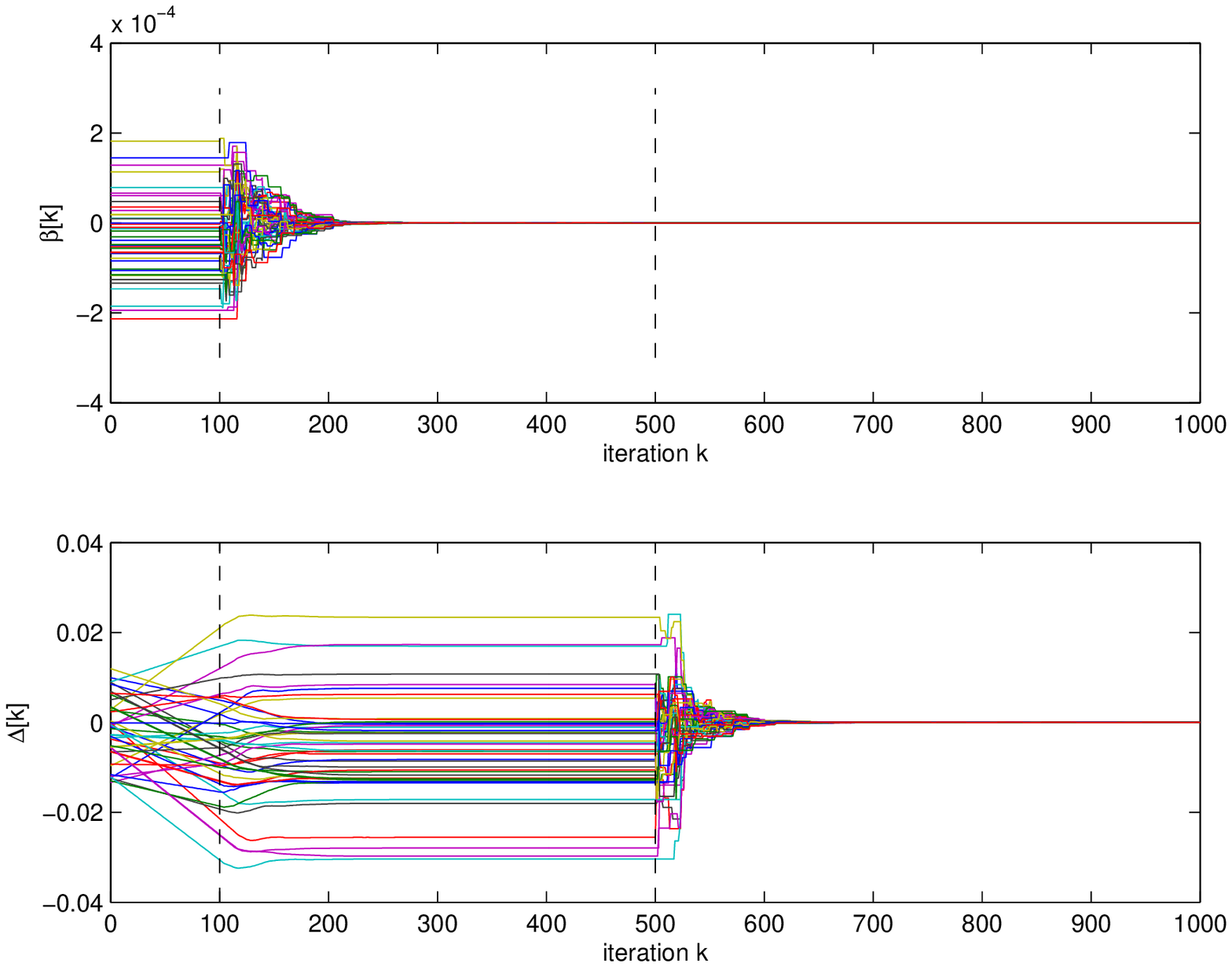}{$N=10$ node synchronization example with equiprobable transmit/receive pairs and fixed \ss $\mu = 0.5$.}{3.45}

\putFig{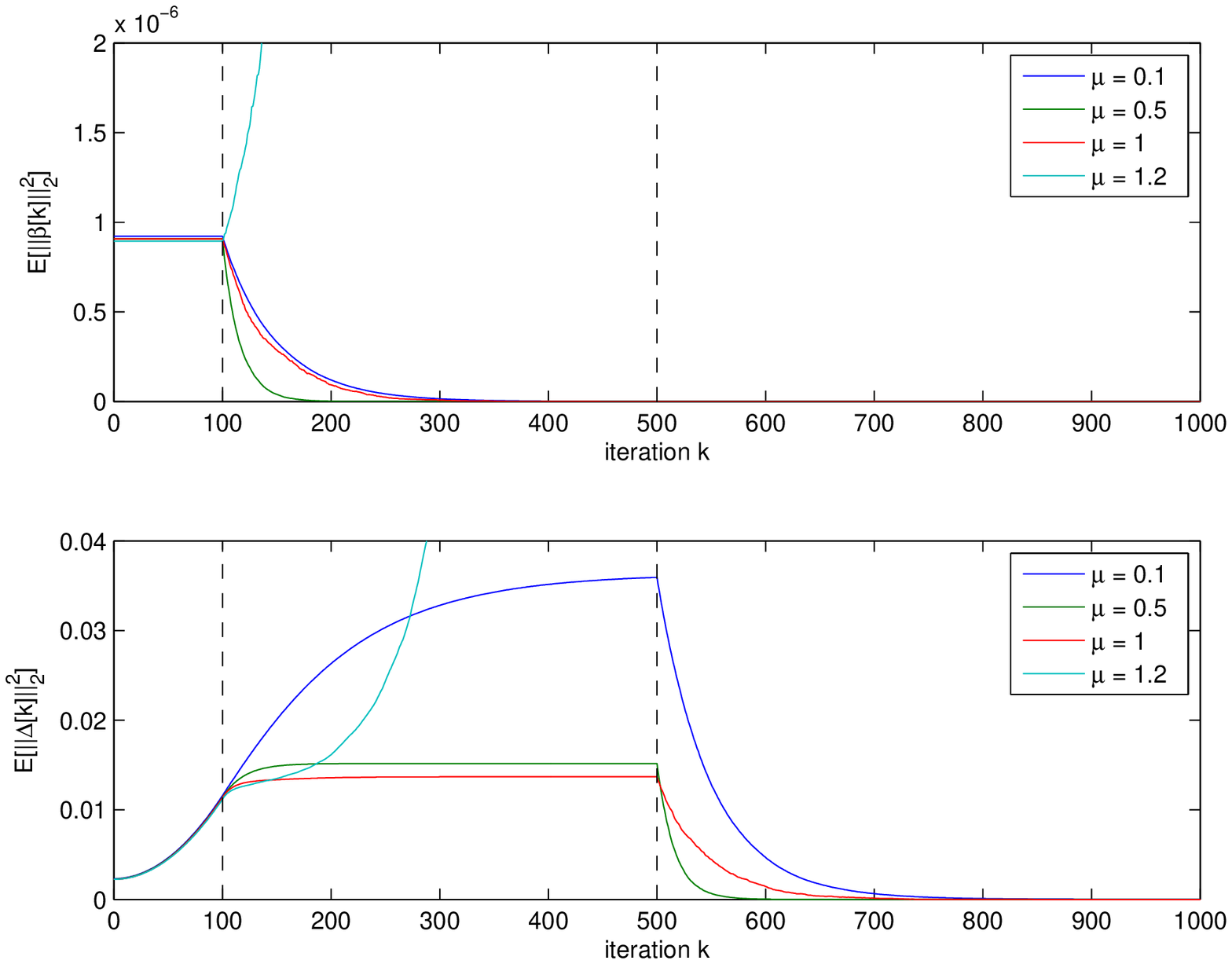}{Empirically averaged pairwise drift and pairwise offset norms for $N=10$ node synchronization with equiprobable transmit/receive pairs.}{3.45}

\subsection{Partitioned Network with Gateway Node}
In this example, the $N=10$ node network is partitioned into sets $\mathcal{S}_1 = \{1,2,3,4,5\}$ and $\mathcal{S}_2=\{5,6,7,8,9,10\}$ where communication between members of the same set has probability $p_{i,j}=\frac{1}{50}$ and communication between members of different sets has probability $p_{i,j}=0$. Figure~\ref{fig:n10_gateway} illustrates this case. Observe that node~5 is the only member of both sets, hence we refer to node~5 as the ``gateway node'' even though node~5 has no special functionality. From Theorem~\ref{thm:generalcase}, we can numerically determine the upper bound on the \ss to be $\mu < 1.11$.

\putFig{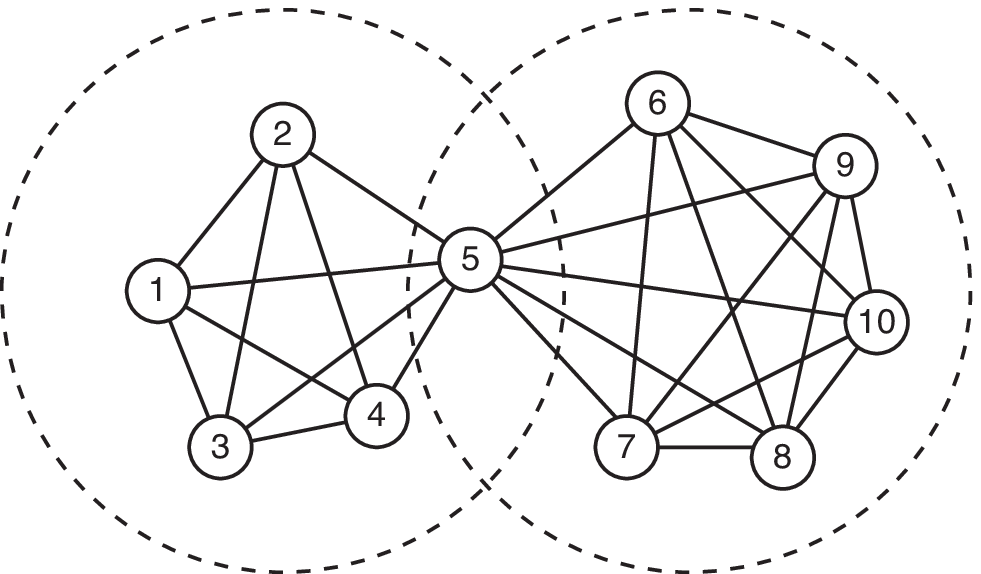}{$N=10$ partitioned network topology with node~5 serving as the gateway node. Each edge in the graph corresponds to a sender/receiver pair with probability $p=\frac{1}{50}$ in either direction.}{2.5}

Figure~\ref{fig:n10_gateway_singlerealization} shows a single realization of the pairwise drift $\vec{\beta}[k]$ and pairwise offset $\vec{\Delta}[k]$ processes for $k=0,\dots,1000$ and fixed \ss $\mu=0.5$. As in the equiprobable case, the pairwise drifts and offsets converge to very small values within a few hundred iterations of the respective compensation algorithms, although convergence is somewhat slower in this example when compared to Figure~\ref{fig:n10_equiprobable_singlerealization}.

Figure~\ref{fig:n10_gateway_averagenorms_multimu} shows the pairwise synchronization metrics $\| \vec{\beta}[k] \|^2_2$ and $\| \vec{\Delta}[k] \|^2_2$, averaged over 1000 Monte-Carlo realizations. We again see divergence when $\mu=1.2>1.11$ and monotonic expected convergence for $\mu = 0.1,0.5,1$. The main difference in this example with respect to the equiprobable case is that the convergence tends to be slower since the two sets only exchange messages through node~5. In particular, in the $\mu=0.1$ case, the convergence is so slow that the drifts are not insignificant before offset compensation begins.

\putFig{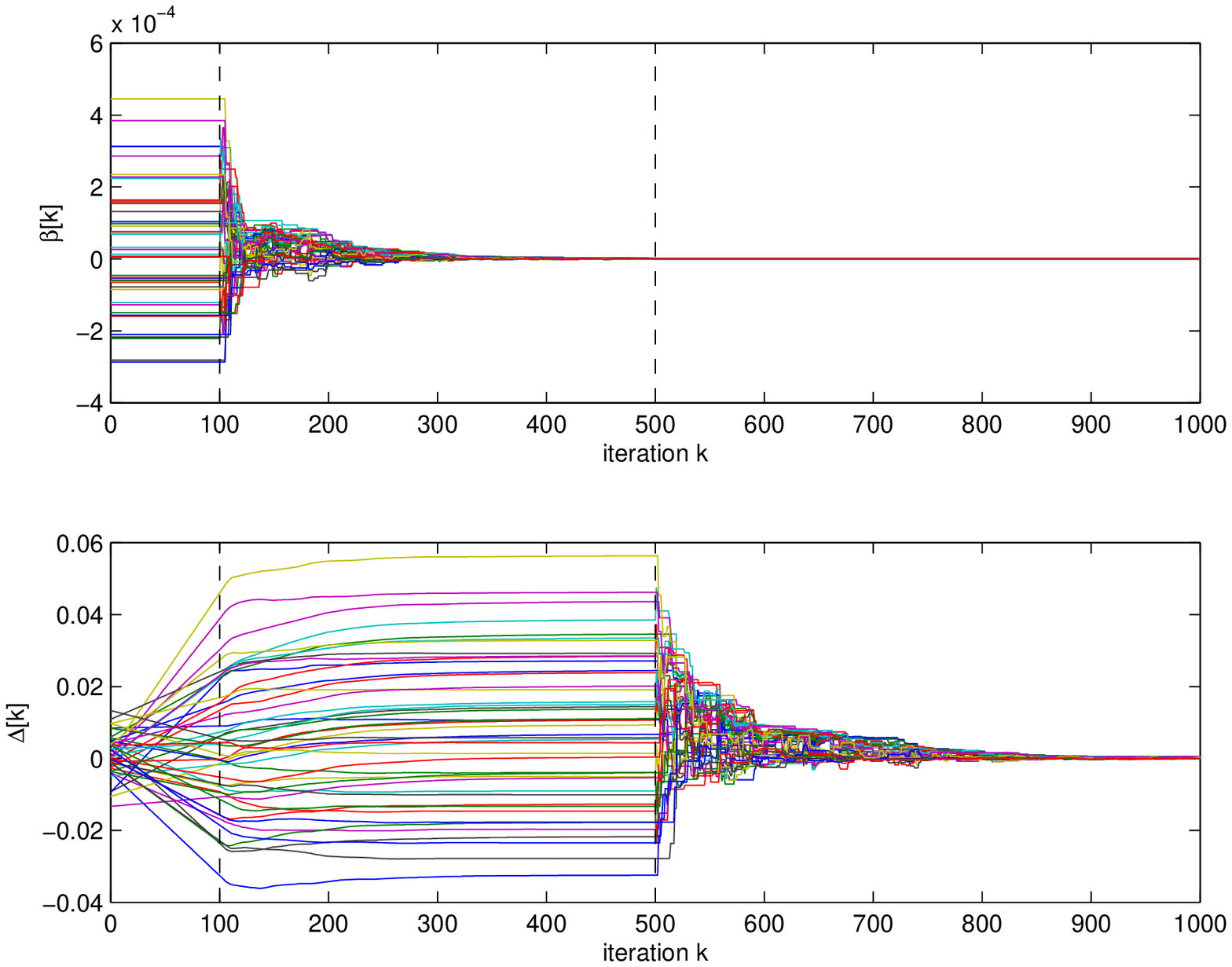}{$N=10$ node synchronization example with partitioned network with gateway node and fixed \ss $\mu = 0.5$.}{3.45}

\putFig{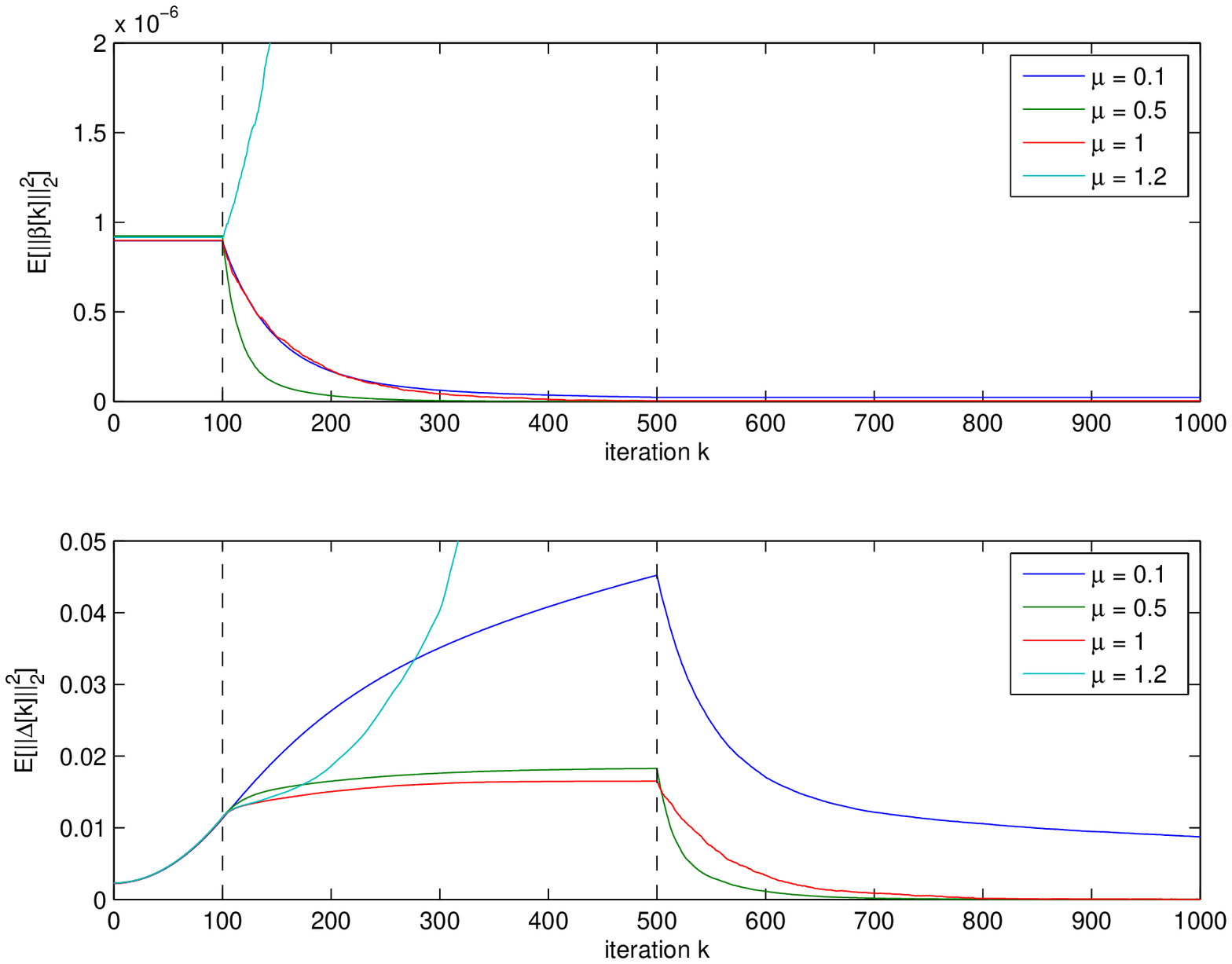}{Empirically averaged pairwise drift and pairwise offset norms for $N=10$ node synchronization with partitioned network with gateway node.}{3.45}

\section{Conclusions and Extensions}
\label{sec:conclusions}
This paper presents a non-hierarchical approach to time synchronization in wireless sensor networks that has  low overhead and can be implemented at the physical and/or MAC layers. Our approach is based on random pairwise bidirectional messaging and allows all nodes to use exactly the same distributed algorithm without requiring local averaging of measurements from other nodes. Analytical results show that the non-hierarchical approach can provide monotonic expected convergence of both drifts and offsets under broad conditions on the network topology and local clock update stepsize. Numerical results are also presented verifying the analysis under two particular network topologies.

Potential extensions of this work include: (i) convergence analysis for simultaneous drift and offset compensation, (ii) the development of explicit bounds on the \ss $\mu$ for network topologies other than equiprobable, and (iii) analysis and simulation of non-hierarchical synchronization with stochastic clocks and/or drift and offset estimations errors.

\appendices

\section{Properties of $\vec{Q}_N$}
\label{app:qproperties}
This appendix provides a recursive definition and several relevant properties of the matrix $\vec{Q}_N$ implicitly defined in (\ref{eq:qimplicit}).  Let $\vec{1}_{N} \in \mathbb{R}^{N}$ be a column vector of all ones.  For $N>2$, $\vec{Q}_N$ has the recursive definition
\begin{eqnarray*}
\vec{Q}_N&=&\left[\begin{array}{c}\begin{array}{c|c}\vec{Q}_{N-1} & \vec{q}_{N-1}\end{array}\\\hline
\vec{I}_{N-1}
\end{array}\right]
\end{eqnarray*}
where $\vec{Q}_2=1$.  Recalling $\vec{q}_N = [0,\dots,0,-1,\dots,-1]^\top \in \mathbb{R}^{N(N-1)/2}$ with the last $N-1$ elements of $\vec{q}_N$ equal to $-1$, we note that
\begin{eqnarray}
\vec{Q}_N^\top\vec{q}_N&=&-\vec{1}_{N-1}.
\label{eq:oldQ4}
\end{eqnarray}
Hence,
\begin{eqnarray}
\vec{Q}_N^\top\vec{Q}_N
&=&
\left[\begin{matrix}\vec{Q}_{N-1}^\top\vec{Q}_{N-1}+\vec{I}_{N-2} & -\vec{1}_{N-2} \\
-\vec{1}_{N-2}^\top & N-1\end{matrix}\right]\nonumber\\
&=&N\vec{I}_{N-1}-\vec{1}_{N-1}\vec{1}_{N-1}^\top\label{eq:QQ}
\end{eqnarray}
where the last equality can be shown by induction.

Furthermore, the product $\vec{Q}_N^\top \vec{Q}_N$ has $N-2$ eigenvalues equal to $N$ and one eigenvalue equal to one. To see this, define
\[
\vec{V} = \left[\begin{matrix}\vec{I}_{N-2} & -\vec{1}_{N-2}\\
-\vec{1}^\top_{N-2} & -1\end{matrix}\right] \in \mathbb{R}^{(N-1) \times (N-1)}
\]
where $\vec{V}$ is full rank. By the matrix inversion lemma \cite{Horn:Book:94}, we can write
\[
\vec{V}^{-1}=\frac{1}{N-1}\left[\begin{matrix}(N-1)\vec{I}_{N-2}-\vec{1}_{N-2}\vec{1}^\top_{N-2} & -\vec{1}_{N-2}\\
-\vec{1}^\top_{N-2} & -1\end{matrix}\right]
\]
so that indeed $\vec{VV}^{-1}=\vec{I}_{N-1}$.

Recall that $\ee_{N-1} \in \mathbb{R}^{N-1}$ is a column vector of zeros with a one in the last position.  Then $\vec{V}(\ee_{N-1}\ee_{N-1}^\top)\vec{V}^{-1}$ is the outer product of the last column of $\vec{V}$ and the last row of $\vec{V}^{-1}$, hence
\begin{equation*}
  \vec{V}(\ee_{N-1}\ee_{N-1}^\top)\vec{V}^{-1}=\frac{1}{N-1}\vec{1}_{N-1}\vec{1}_{N-1}^\top.
\end{equation*}
Consequently, we can write
\begin{align*}
\vec{Q}_N^\top \vec{Q}_N &={} N \vec{I}_{N-1} - \vec{1}_{N-1}\vec{1}^\top_{N-1} \\
&={} N\vec{VV}^{-1}-(N-1)\vec{V}(\ee_{N-1}\ee_{N-1}^\top)\vec{V}^{-1}\\
&={}\vec{V}\left(N\vec{I}_{N-1}-(N-1)\ee_{N-1}\ee_{N-1}^\top\right)\vec{V}^{-1}
\end{align*}
where the diagonal matrix
\begin{equation}
  N\vec{I}_{N-1}-(N-1)\ee_{N-1}\ee_{N-1}^\top = {\rm diag}(N,\dots,N,1)
  \label{eq:Qevals}
\end{equation}
contains the eigenvalues of $\vec{Q}_N^\top\vec{Q}_N$.

\section{Proof of Theorem 1}
\label{app:generalcase}
\begin{IEEEproof}
Given the pairwise drift vector $\vec{\beta}[k]$ as defined in (\ref{eq:beta}) and pairwise drift update vectors $\vec{b}_{i,j}[k]$ as defined in (\ref{eq:driftupdate}) with \ss $\mu$. For monotonic expected convergence, we will show $\E\left\{ \| \vec{\beta}[k+1] \|_2^2 \, | \, \vec{\beta}[k]\right\} < \| \vec{\beta}[k] \|_2^2$ for all $\vec{\beta}[k] \in {\rm range}(\vec{Q}_N)$ satisfying $  \vec{\beta}[k] \neq 0 $ where the expectation is taken over all $N(N-1)$ possible sender/receiver pairs. Since
\begin{align*}
  \| \vec{\beta}[k+1] \|_2^2 &={} \| \vec{\beta}[k] + \mu \vec{b}_{i,j}[k] \|_2^2 
\end{align*}
and $\vec{\beta}[k] = \vec{Q}_N\vec{\beta}_N[k]$, we have monotonic expected convergence if and only if
\begin{equation*}
  2\mu\vec{\beta}^\top_N[k] \vec{Q}_N^\top \E\{\vec{b}_{i,j}[k] \, | \, \vec{\beta}[k]\} + \mu^2\E \left\{ \| \vec{b}_{i,j}[k] \|_2^2 \, | \, \vec{\beta}[k] \right\} < 0.
\end{equation*}
From (\ref{eq:R}) and (\ref{eq:S}), this condition can be be equivalently expressed as
\begin{equation}
  \label{eq:condition2}
  \vec{\beta}^\top_N[k] \vec{Q}_N^\top \left(\mu\vec{S}-2\vec{R}\right)\vec{\beta}[k] \vec{Q}_N\vec{\beta}_N[k]  < 0.
\end{equation}
Note that $\vec{R}$ is not necessarily symmetric. Hence, (\ref{eq:condition2}) is true for all $\vec{\beta}_N[k] \neq 0$ if and only if
\begin{align*}
 \vec{A} &={} \vec{Q}_N^\top \left( \frac{(2 \vec{R}-\mu \vec{S})+(2 \vec{R}-\mu \vec{S})^\top}{2}\right)\vec{Q}_N \\
 & ={} \vec{Q}_N^\top(\vec{R}+\vec{R}^\top - \mu\vec{S})\vec{Q}_N 
\end{align*}
is positive definite.
\end{IEEEproof}

\section{Proof of Theorem 2}
\label{app:equiprobable}
\begin{IEEEproof}
Using the notation and definitions from Appendix~\ref{app:generalcase} with the additional assumption of equiprobable sender/receiver pairs such that $p_{i,j} \equiv p = \frac{1}{N(N-1)}$ for all $i,j$, we will show that
\begin{align*}
 \vec{A} &={}  \vec{Q}_N^\top(\vec{R}+\vec{R}^\top - \mu\vec{S})\vec{Q}_N 
\end{align*}
is positive definite if and only if $\mu < \frac{N}{N-1}$. Our approach is to show $\vec{A} = f(\mu) \vec{Q}_N^\top \vec{Q}_N$ under the equiprobable assumption and to use properties of the $\vec{Q}_N$ matrix from Appendix~\ref{app:qproperties} to conclude that $\vec{A}$ is positive definite under this assumption if and only if $f(\mu)>0$.

Observe that entry $(m(i,j),i)$ of $\bar{\vec{Q}}_N$ is equal to $+1$ and entry $(m(i,j),j)$ of $\bar{\vec{Q}}_N$ is equal to $-1$, for all $1 \le i < j \le N$, with all other entries of $\bar{\vec{Q}}_N $ equal to zero. Hence
\begin{equation*}
  \bar{\vec{Q}}_N = \left(\sum_{i=1}^N \sum_{j \neq i} \vec{E}_{i,j}\right)^\top  
\end{equation*}
where $\vec{E}_{i,j} \in \mathbb{R}^{N \times N(N-1)/2}$ is a matrix of all zeros except for a ${\rm sign}(j-i)$ in position $(i,m(i,j))$ if $i<j$ or position $(i,m(j,i))$ if $i>j$.
Under the equiprobable assumption, $\vec{R}$ can be written as
\begin{align*}
  \vec{R} &={} \frac{1}{N(N-1)}\bar{\vec{Q}}_N  \sum_{i=1}^N \sum_{j \neq i} \vec{E}_{i,j}  \\
          &={} \frac{1}{N(N-1)}\bar{\vec{Q}}_N  (\bar{\vec{Q}}_N)^\top \\
          &={} \frac{1}{N(N-1)} \left(\vec{Q}_N \vec{Q}_N^\top + \vec{q}_N \vec{q}_N^\top\right).
\end{align*}
Note that $\vec{R}$ is symmetric in this case, hence $\vec{R}+\vec{R}^\top = 2\vec{R}$. We can further compute
\begin{align*}
  2\vec{Q}_N^\top \vec{R} \vec{Q}_N &={} \frac{2}{N(N-1)} \left(\vec{Q}_N^\top \vec{Q}_N\vec{Q}_N^\top \vec{Q}_N \! + \! \vec{Q}_N^\top\vec{q}_N \vec{q}_N^\top \vec{Q}_N \right) \\
    &={} \frac{2}{N(N-1)}  \Bigl( (N \vec{I}_{N-1} - \vec{1}_{N-1}\vec{1}^\top_{N-1})^2 + \\
    & \qquad (-\vec{1}_{N-1}) (-\vec{1}^\top_{N-1}) \Bigr) \\
    &={} \frac{2}{N(N-1)} \Bigl( N^2 \vec{I}_{N-1} -N\vec{1}_{N-1}\vec{1}^\top_{N-1}  \Bigr) \\
    &={} \frac{2}{N-1} \vec{Q}_N^\top \vec{Q}_N
\end{align*}
where we used (\ref{eq:oldQ4}) and (\ref{eq:QQ}) from Appendix \ref{app:qproperties} in the second equality and (\ref{eq:QQ}) again in the final equality.

In the equiprobable case we also have $\vec{S} = 2p(N-1)\vec{I}_{N-1}$, hence
\begin{align*}
  \vec{Q}_N^\top \vec{SQ}_N &={} \frac{2}{N} \vec{Q}_N^\top \vec{Q}_N.
\end{align*}
It follows that
\begin{equation*}
  \vec{Q}_N^\top(\vec{R}+\vec{R}^\top - \mu\vec{S})\vec{Q}_N = \left(\frac{2}{N-1} - \mu \frac{2}{N}  \right) \vec{Q}_N^\top \vec{Q}_N
\end{equation*}
which, from the fact that the eigenvalues of $\vec{Q}_N^\top \vec{Q}_N$ are strictly positive as shown in (\ref{eq:Qevals}), is positive definite if and only if $\mu < \frac{N}{N-1}$.
\end{IEEEproof}

\bibliographystyle{IEEEtran}
\bibliography{IEEEabrv,sync}

\clearpage
\end{document}